# Physical theory of the twentieth century and contemporary philosophy

Miloš V. Lokajíček, Institute of Physics, AVČR, 18221 Prague, Czech Republic

**Abstract**

It has been shown that the criticism of Pauli as well as of Susskind and Glogover may be avoided if the standard quantum-mechanical mathematical model has been suitably extended. There is not more any reason for Einstein's criticism, either, if in addition to, some new results concerning Bell's inequalities and Belifante's argument are taken into account. The ensemble interpretation of quantum mechanics (or the hidden-variable theory) should be preferred, which is also supported by the already published results of experiments with three polarizers. Greater space in the text has been devoted also to the discussion of epistemological problems and some philosophical consequences.

**1. Introduction**

One may say that Aristotle (383-321 b. C.) started with learning to know systematically the world. He put the foundations not only to natural sciences, but also to metaphysics when he tried to derive also some further consequences following from the ontological nature of matter. There is a close coherency between these two regions, since the metaphysical knowledge should be in the full harmony with physical (resp. scientific) knowledge, and also vice versa, as will be seen in the following.

The Aristotelian approach was not applied to in the Europe practically in the whole first thousand years a. C. It was again discovered for Europe and further developed by Thomas Aqu. (1225-74), who was the follower of Albertus Magnus (1193-1280), being denoted as the greatest scientist at that time. The activities of Thomas involved, however, also the whole region of theological philosophy that started from the Christian revelation at the difference to metaphysical considerations being based only on world observation. Thomas was able, of course, to distinguish in principle between theses two approaches.

It seems, however, that at least some followers of Thomas came to statements in which the two mentioned approaches were not sufficiently distinguished or some metaphysical conclusions did not respect new pieces of knowledge following from world observation. It was probably main reason, why in the new age a series of different philosophical doctrines appeared, from which it was started also when new naturally scientific theories (hypotheses) were formulated. Any metaphysical consideration (based on ontological characteristics) was abandoned. Ernst Mach (1838-1916) in the end of the 19th century (see, e.g., [1]) refused any metaphysics, which influenced the whole further development of physical science in decisive way.

The physics limited to quite phenomenological considerations and phenomenological models of natural phenomena and did not take into account the matter reality of natural objects and natural facts. The world picture was gradually formed that consisted before all in mathematical constructions being derived in some way from the observation of natural phenomena, or rather from some set of measured values. And one is entitled to say that on such a basis both the main theories of the 20$^{th}$ century were formulated: quantum mechanics as well as special relativity theory. In the following we will deal to a greater detail with the quantum mechanics, as we have finished recently corresponding analyses.

It is necessary to start with more general consideration (see Sec. 2), what role is played in these approaches by the capability of human reason to get to know the nature and to draw conclusions from world observation, which is the only source of scientific (as well as metaphysical) knowledge; the logical rules that represent an indivisible part of the whole knowledge process will be also mentioned. In Sec. 3 the mathematical models of natural processes will be discussed, and also the difference between phenomenological and ontological models. Sec. 4 will be devoted to the basic assumptions of quantum mechanics and to two critical comments concerning the corresponding mathematical model and being discussed in literature all the time. Einstein's criticism being discussed practically also the whole century will be handled in Sec. 5. Sec. 6 will be devoted to the results of EPR experiments and to the meaning of Bell inequalities that represent the basis of all contemporary experimental and theoretical analyses. The transmission of the light through two polarizers and actual meaning of corresponding EPR experiments will be handled in Sec. 7. The results of experiments with three polarizers that bring a quite new insight into the whole situation will be introduced in Sec. 8. The importance of ontological models for metaphysical consideration will be mentioned in Sec. 9.

The goal of the presented paper consists in summarizing some new results that have been obtained by the author with the help of other collaborators and in introducing some new points of view. It contains also the standpoint of a natural scientist to some interdisciplinary problems that were subjects of his interest during his professional life. There have been, however, some difficulties with presenting the corresponding new results. On one side it has been necessary to express individual statements relatively shortly, but on the other side sufficiently exactly. Some statements will be, therefore, fully understandable at the first reading to specialists from a given region only. In any case the reader must access to all statements with some reliance before he will be able to convince himself about their rightfulness on the basis of other papers quoted in the end of this text.

**2. The capability of human reason and the knowledge of world**

All world knowledge of ours is based on the observation of natural phenomena with the help of our senses and at the present also on the results of detailed measuring approaches. On the basis of observation and measurement and with the help of our reason (using logical induction, resp. on the basis of intuition) we are trying to formulate some more general statements or unifying hypotheses. The goal of further approaches and analyses consists then in deriving all possible consequences that follow from a given hypothesis (or a set of hypotheses). As such hypotheses start always from a limited set of our pieces of knowledge it is clear immediately in some cases that they cannot sustain in further considerations. Generally, it is then necessary to derive for any hypothesis all possible consequences with the help of logical deduction and to analyze, whether some logical contradictions do not exist, and further to compare the given consequences to all possible observations of natural phenomena. If a contradiction is found the given hypothesis must be refused, resp. modified, so as the given contradiction might be removed. If one does not come after sufficient falsifying effort to any contradiction the given hypothesis (or the set of hypotheses) may be denoted as plausible; and it is possible to start from it in further considerations about the natural laws concerning the world and human being.

It follows from the preceding that the falsification approach represents important and practically basic knowledge method based on human reason, as K. Popper (1902-94) proposed in the first half of the 20$^{th}$ century. One is never justified to denote our reason knowledge for a verified truth, as one can never grant that in following steps a logical contradiction or a contradiction to nature observation will not appear. On the other side one must accept any non-falsified hypothesis as plausible (and in principle equivalently valid), even if it is in contradiction to another plausible hypothesis. The decision concerning the preference of one of them must be left to other logical and experimental analyses.

The falsification approach concerns, of course, only our possibilities of getting to know the truth about the world; the existence of the truth is not being refused. Even the certain knowledge of truth may exist which consists in the set of negative (falsification) statements that were obtained by falsification approaches in the past. However, any positive statement (i. e., a hypothesis) cannot be denoted as certain truth. Each statement of such a kind that was not falsified may be regarded as plausible with that it may be falsified in the future.

In the course of the 20th century the simple falsification approach was transformed in the scientific world (before all in the region of physics) to the ideology of falsifiability. Natural falsification approach was abandoned in principle and any phenomenological mathematical model (i.e., a hypothesis) was denoted as verified if it described a given set of experimental (measured) data in an acceptable way. However, on the basis of falsifiability principle any other (even quite plausible) model was refused as unnecessary if it did not lead to some predictions differing from those of the preceding model. In many cases the way to actually true world knowledge was thus closed.

Our knowledge does not end by natural science only. We are entitled (and in principle obliged) to formulate also the hypotheses that start not only from the direct observation and measurement, but also from ontological considerations concerning matter world. Even in such a case we should analyze all possible consequences with the help of logical deduction approaches and denote as falsified the hypotheses that lead to contradiction with natural phenomena. On the other side the hypotheses for which any contradiction has not been found should be denoted as plausible. I.e., the solution of metaphysical problems should continue, as based by Aristotle and developed further by Thomas Aqu. It does not mean, of course, that we should accept all previous metaphysical statements as dogmas; it is necessary to access to them equally as to scientific hypotheses (resp. models or theories) and to continue always in falsifying approach.

They are also the logical rules that must be denoted as an indivisible part of deduction approaches. Since the time of Aristotle they start from the two-value logic. They represent one of basic hypotheses in the region of scientific knowledge that might be falsified in principle as other scientific hypotheses, which did not happen, of course, until now. In relation to Copenhagen quantum mechanics the more-value logics were discussed in the past century. However, nobody succeeded in applying such logic to natural phenomena. We shall see in the following that no reason exists for abandoning the two-value logic in the future, either. It represents the basic metaphysical hypothesis that is a part of all natural sciences. If one did not start from such a basis it would not be possible to speak about any science at all.

**3. Mathematical modeling in contemporary natural science**

Contemporary scientific (and mainly physical) research would be practically impossible if corresponding mathematical models had not been made use of. They were developed mainly by G. Galilei (1564-1642) and I. Newton (1643-1727). Their models must be denoted as ontological as they started from ontological ideas about the matter world.

However, later the preference was given to mathematical models that represented certain sets of measured values without taking any interest about the matter mechanism being hidden behind. Such models that may be denoted as phenomenological were supported in decisive way by positivistic philosophy that denoted measured values as the only positive facts. At the present mainly these phenomenological models are made use of in the natural sciences.

Also both the main physical theories of the 20th century, i.e. quantum mechanics and special relativity theory, belong to the class of phenomenological models, as they have started in principle from mathematical assumptions and rules only. It is hardly possible to attribute to them an ontological model that would be in sufficient agreement with general experience concerning matter world. On their basis it is rather argued against the ontological nature of the world, without bringing corresponding conclusions to a sufficient agreement with current experience.

The great discoveries at the break of the 19th and 20th centuries contributed to the given situation in significant way; they brought a new look at phenomena at the level of microscopic world (discovery of X and gamma radiations and of electron (1985-7), and further, quantum energy transfer (1900) and photon prediction (1904)). A broad space was opened for formulation of new hypotheses that could not be tested directly, but only on the basis of indirect effects with the help of macroscopic measuring devices. And in such a situation the phenomenological models seemed to be advantageous and both the mentioned theories were proposed.

In the following we shall deal to a greater detail, as already mentioned, with one of these theories, i.e., with the quantum mechanics in its Copenhagen interpretation, since practically all authors relate to this alternative in describing world behavior. In experimental tests and in their interpretations they do not take, however, always into account all assumptions, on which Copenhagen alternative is based. In many cases already the agreement with predictions of Schroedinger equation (i.e. with one of the assumptions) is denoted as agreement with the whole Copenhagen quantum mechanics. This alternative includes, of course, some further important assumptions that are not in full agreement with Schroedinger equation. In the following the corresponding results (including our analyses obtained in the recent time) will be summarized.

### 4. Quantum mechanics – assumptions and basic critical comments

It is necessary to start with introducing the assumptions the quantum mechanics is based on. We shall limit ourselves to the main of them and avoid the details that are rather of mathematical and technical kinds. Then it is possible to say that the quantum mechanics is based on three main assumptions:

 (i) Time evolution of any physical system may be characterized by the wave function $\psi(x,t)$ where $t$ is time parameter and $x$ represents space coordinates of all particles the given system consists of. This wave function must fulfill the linear differential equation

$$i\hbar \frac{\partial}{\partial t}\psi(x,t) = H\psi(x,t)$$

denoted as time-dependent Schroedinger equation, the basis of its being formed by the Hamiltonian

$$H = \sum_j \frac{p_j^2}{2m} + V(x) \ ,$$

i.e., by the expression characterizing total energy of the corresponding system, given as the sum of kinetic and potential energies. Kinetic energy is then given by the sizes of particle momentums, the components of which are substituted in the Schroedinger equation by derivatives according to corresponding coordinates (multiplied by $\hbar/i$). The values of all physical characteristics at any $t$ may be then derived from the wave function according to strict rules (as the expected values of corresponding operators). If one limits to the basic solutions (containing always only one eigenfunction of the corresponding Hamiltonian) these values correspond to those for individual particles as in the classical physics. One obtains alternative description of corresponding classical behavior [2]. However, two different interpretation alternatives of Schroedinger equation might exist (at east in principle) in the case of superposition solutions: Copenhagen (orthodox) and statistical (ensemble). According to the latter the given solution should represent a statistical set of different systems described by basic solutions, while according to the former again one particle should be represented, no analogue existing for its properties in the classical world. And just to this former alternative the following two currently used assumptions correspond.

 (ii) The second assumption defines the corresponding mathematical Hilbert space in which the solutions of any Schroedinger equation may be represented. It should be spanned on one set of eigenfunctions of corresponding Hamiltonian. It means that, e.g., the different physical contents of Schroedinger solutions evolving in opposite time directions are not practically distinguished.

 (iii) Copenhagen interpretation involves then yet further assumption, according to which the mathematical superposition principle (holding in any metric space) is interpreted as important physical characteristics. The mathematical superposition relations between different vectors are interpreted as actual relations between corresponding physical states. It means that according to this assumption also other (from classical point of view quite different) states may be always contained with some probabilities in the state represented by one vector in the Hilbert space.

The mathematically defined model has represented, therefore, the decisive criterion for physical interpretation; any attention has not been devoted to matter mechanism (ontological nature). However, already in 1933 W. Pauli [3] called attention to the fact that the given system of assumptions led to a problem in the description of time evolution. Schroedinger equation has described uniquely the whole time evolution; and Pauli showed that in the given Hilbert space (fulfilling the second assumption) the time evolution might be represented ordinarily by a time operator only if the Hamiltonian possessed continuous spectrum of all real values, i.e., from $-\infty$ to $+\infty$. That contradicts, of course, physical requirements, as the energy (represented by Hamiltonian) cannot be negative (going to infinity). In the Hilbert space spanned on one set of Hamiltonian eigenfunctions the time evolution must be interpreted in the way where no causal evolution is possible and the probability (or chance) plays a main role.

Further criticism of the given mathematical system was added in 1964 by Susskind and Glogover [4]. They showed that the exponential time operator did not fulfill the requirement of unitarity for all vectors of the given Hilbert space; i.e. the probability has not been conserved in its framework. The problem is important before all, when the corresponding Hamiltonian has a discrete spectrum.

In the past years many papers were published that tried to remove these deficiencies. However, all attempts of solving both the critical remarks as one common problem were unsuccessful. It is possible to say that the solution was found only recently when it was shown that the two mentioned deficiencies may be removed in two subsequent steps, i.e. when the standard Hilbert space is twice doubled as shown in papers [5,6]. Both these steps were proposed in principle in 1967, by Lax and Phillips [7] and by Fain [8]. However, the meanings of these proposals were not well understood earlier.

The given extension of Hilbert space has, of course, an important consequence. It is necessary to change correspondingly the second quantum-mechanical assumption (i.e., to extend the Hilbert space) and to abandon fully the third one. Some important physical consequences follow from it, even if the proper mathematical model changes only unsubstantially, which will be mentioned later. Now we will try to show, that the proposed modification (extension) of quantum-mechanical model is supported decisively by more important (rather experimental) arguments that have been based on the critical analysis having been presented by Einstein and collaborators [9] in 1935.

### 5. The criticism of Einstein

Critical comments discussed in the preceding section that have been removed by the extension of Hilbert space have concerned internal discrepancies in the mathematical model based on the mentioned assumptions. A. Einstein was bringing important objections against the Copenhagen quantum mechanics also from the experimental point of view from the very beginning. However, only in 1935 he succeeded in their full formulation. It occurred too late, when the further development of the given physical theory could not be practically influenced.

The problem concerned in principle the existence of the so called hidden parameters (variables). On the basis of the second assumption the set of parameters distinguishing individual states at any time was reduced in comparison with classical physics. And it was discussed whether corresponding additional parameters (denoted as hidden) have physical meaning in the microworld or not. The first answer was given by J. von Neumann [10] in 1932, who derived on the basis of a special assumption that there is not any space for them. And physical community in its majority has accepted Bohr's Copenhagen quantum mechanics as the decisive theory of microworld.

Einstein [9] proposed, however, a Gedankenexperiment known in literature as EPR experiment (according to all co-authors). He demonstrated with its help that under certain conditions Bohr's theory required for two distant measuring devices to influence mutually one another. He assumed that two particles formed by the decay of one particle in the rest and running in opposite directions were detected. If one measures coincidence probabilities of both the particles in two opposite places the result in one measuring device depends according to Copenhagen quantum mechanics on the result obtained earlier by the other measuring device.

Bohr [11] refused Einstein's criticism and expressed his standpoint that the microworld exhibited the given property (paradoxical according to macroscopic experience). The assumption was formulated that microscopic particles were not locally limited and that they might exhibit their presence and their effects even on macroscopic distances. The physical community accepted such an argumentation even if Einstein did not abandon his criticism till the end of his life.

As to the description in the framework of the given assumptions and of the given mathematical model the problem consisted in the question, whether each vector of the given Hilbert space (fulfilling the second assumption in Sec. 4) represented a physical state. Or from the other side, whether individual states are distinguished only mathematically by Hilbert space vectors or also with the help of the mentioned hidden parameters. Or, whether all three assumptions from Sec. 4 are valid in agreement with the then conviction.

Common opinion of physical community changed partially in 1952 when D. Bohm [12] showed that a certain hidden parameter existed already in Schroedinger equation. And a greater change occurred in 1964 when J. Bell [13] performed a more detailed analysis from which it followed that the argument presented by von Neumann to support Copenhagen quantum mechanics started from an unphysical assumption. According to Bell there was not any theoretical reason against the existence of hidden parameters, and therefore, any decisive reason for the Copenhagen interpretation of quantum mechanics. The question was shifted to the stage when it was fully opened from the theoretical point of view and could be answered only on experimental basis. It should have been decided with the help of some inequalities that were derived also by Bell; they should be valid in the case if the hidden parameters existed and violated in the experiments of EPR type in the case of Copenhagen interpretation. Many authors tried then to work out Einstein's Gedankenexperiment to be experimentally feasible. They succeeded and corresponding experiments were performed in years 1971-82.

**6. Bell's inequalities and the results of EPR experiments**

The experiments actually performed differed partially from the Gedankenexperiment proposed originally by Einstein; instead of the mere detection of particles the spin orientations of two photons were measured. Two light photons were emitted from an excited atom that was in the rest and the spin of which equaled zero; they were running in opposite directions:

$$^{\alpha}|\leftarrow---\circ---\rightarrow|^{\beta}$$

Both the photons had the same polarization (opposite spins); the polarization of individual pairs was, however, quite random. The measurement consisted in determining coincidence probabilities for the transmissions of both the photons through two polarizers at different orientations of their axes, deviated by angles $\alpha$ and $\beta$ from the common zero position.

As already mentioned Bell [13] derived the inequalities holding for individual transmission probabilities of equally polarized photons at different deviations of polarizer axes. It should hold for any four different coincidence measurements

$$B = p_1(\alpha)p_2(\beta) + p_1(\alpha)p_2(\beta') + p_1(\alpha')p_2(\beta) - p_1(\alpha')p_2(\beta') \leq 2$$

where $p_1(\alpha)$ and $p_2(\beta)$ are probabilities of photon transfer through one (or the other) polarizer at angle deviations $\alpha$ and $\beta$.

And the goal of corresponding EPR experiments consisted in establishing, whether the given inequality was violated for some combinations of angles $\alpha$ and $\beta$ or not. Its violation should have corresponded to the validity of Copenhagen interpretation, while in the opposite case the preference should be given to the theory with hidden parameters. The experiments were closed in 1982 with the following result (see [14]): It was concluded that Bell inequalities were violated and that the given measurements were practically in agreement with the predictions of the quantum mechanics (which was derived practicaly from Schroedinger equation only, without having taken other assumptions into account).

That was regarded at that time as decisive confirmation of Copenhagen quantum mechanics. The given conclusion was significantly supported also by the argument provided in 1973 by Belinfante [15]. He argued that the predictions of quantum mechanics and of any theory with hidden parameters must be significantly different. Thus, the agreement with quantum mechanics was regarded as a further confirmation that the theory with hidden parameters must be refused on the basis of given experimental data. However, it did not represent the end of discussions about Einstein's criticism since the paradox of non-localized microscopic particles as well as the discrepancies in the mathematical model (discussed in Sec. 4) continued.

In addition to, we have succeeded in showing quite convincingly that none of the two mentioned arguments (i.e., Bell's inequalities and Belinfante's statement) can be applied reasonably in the framework of complete hidden-parameter theory to the given experiment. As to the inequalities Bell attributed certain polarization to a given photon pair, but he did not take the particle character of individual photons into consideration at all. In such a case it was not possible to interchange arbitrarily quantities $p_1(\alpha)$ and $p_1(\alpha')$ (resp. $p_1(\beta)$ and $p_1(\beta')$) corresponding to divers photon pairs, as Bell was forced to do to get the limit of 2.

For the first time we tried to call the attention to this fact in paper [16] where the earlier approaches of deriving Bell's inequalities (summarized in Ref. [17]) were analyzed from this point of view. As to the original approach of Bell it is not possible to pass in the general case (particle character being respected) from Eq. (3.11) to Eq. (3.12) in paper [17]. Without the simplified description of Bell it is not possible to pass from Eq. (3.18) to Eq. (3.19) in the approach proposed by Clauser and Horn, either; and it holds also similarly in further approaches reproduced in paper [17]. More detailed points may be found in [16].

The situation may be explained, however, more convincingly if the results shown in [18] are taken into account. The limit of Bell's combination of coincidence probabilities may be influenced by different assumptions as it may be demonstrated with the help of the method of Bell's operator

$$B = a_1 b_1 + a_2 b_1 + a_1 b_2 - a_2 b_2$$

where $a_j$ and $b_j$ are operators representing corresponding measurements with the help of individual polarizers. According to how different operators commute mutually or do not commute the expected value of the operator $B$ can have three different limits [18]: $2, 2.2^{1/2}, 2.3^{1/2}$, which correspond subsequently to classical description (or to simplified hidden-variable description of Bell), to general theory with hidden parameters (or to statistical interpretation of Schroedinger equation – see [2]) and to Copenhagen interpretation. Bell's limit corresponds, therefore, to the condition considered by von Neumann before years, as it was shown recently also by Malley [19]; however, he derived from this fact improper conclusion introduced in the title of his paper.

It is possible, therefore, to conclude from these results that experimental violation of Bell inequalities does not concern the general theory of hidden variables. It is, of course, also the Copenhagen alternative that has not been refused, either. The corresponding decision must be looked for on a quite different experimental basis.

The theory of hidden parameters might seem, of course, to be refused further on the basis of the statement, formulated by Belinfante [15] for the case of two polarizers. According to him the predictions of quantum mechanics and of a hidden-variable theory should be significantly different. However, the given statement represents a mistaking conclusion, which will be explained now in Sec. 7.

## 7. Experiments with two polarizers and Belinfante's argument

EPR experiments consist in measuring coincidence transmission probabilities of two equally polarized photons through two polarizers in the dependence on deviation $\alpha$ between polarizer axes. According to experimental results this dependence is practically equal to that obtained in the transmission of one photon through two polarizers. From these experiments it is then known that one must apply the generalized Malus law

$$M(\alpha) = (1-\varepsilon)\cos^2\alpha + \varepsilon,$$

where $\varepsilon > 0$.

The quantum mechanics interprets the light (and in principle each photon) as the superposition of two mutually perpendicular polarizations, one of which is transferred fully through a polarizer and the other is absorbed. It leads to the Malus law where $\varepsilon = 0$. For the fact that $\varepsilon > 0$ the given theory has not practically any explanation. It interprets this non-zero parameter as imperfectness of real polarizers in comparison to ideal polarizers, described by the given theory. Its numerical value is being brought to harmony with that only partial transmission or absorption are attributed to individual basic polarizations.

In the theory of hidden variables it is possible to write for the corresponding transmission

$$P_2(\alpha) = \frac{1}{\pi}\int_0^\pi p_1(\lambda)p_2(\lambda-\alpha)d\lambda,$$

where $\lambda$ is the deviation of photon polarization from the axis of the first polarizer and $p_j(\lambda)$ are transmission probabilities through one polarizer at a given polarization deviation from polarizer axis. The last equation holds for two polarizers when they are arranged one after the other (one photon passes both the polarizers) as well as in coincidence arrangement (EPR experiment – see Sec. 6).

When one puts, e.g.,

$$p_1(\lambda) = 1 - \frac{1-\exp(-(a|\lambda|)^e)}{1+c.\exp(-a|\lambda|)^e)}, \quad a=1.95, \ e=3.56, \ c=500,$$

it is possible to obtain

$$P_2(\alpha) \cong M(\alpha).$$

It follows also that $P_2(\alpha)$ is in the harmony with $M(\alpha)$ where $\varepsilon > 0$, as required by experimental data. The dependencies of corresponding quantities ($p_1(\lambda), P_2(\alpha), M(\alpha)$) are graphically represented in Fig. 1 of paper [18]. Belinfante came to his conclusion (see Sec. 6) when he put without any reason and quite arbitrarily $p_j(\lambda) = \cos^2(\lambda)$.

One can, therefore, conclude that in the case of two polarizers it is possible to obtain in hidden-variable theory in principle the same behavior as in quantum mechanics, and to come to agreement with experimental data. However, as the assumptions concerning the proper mechanisms are very different one should expect that the correspondence is quite accidental and that in other arrangements the predictions will be significantly different. And a suitable experiment will be able to decide which alternative should be preferred. Such a difference should exist, e.g., for the light (photon) transmission through three polarizers, which will be shortly dealt in the next section.

### 8. Measurement of light transmission through three polarizers

Let us assume now that the light goes through three polarizers arranged one after the other

$$\circ ---|---|^\alpha ---|^\beta \rightarrow$$

where α and β are axis deviations of the second and third polarizers from the first one. For transmission probability of the light through three (ideal) polarizers it holds in the quantum mechanics

$$P(\alpha,\beta) = \cos^2\alpha\cos^2(\alpha-\beta).$$

As to the theory with hidden parameters we performed the preliminary numerical analysis of corresponding predictions already earlier under the assumption that polarization characteristics of individual polarizers may be described by expressions introduced in Sec. 7. It was possible to show that measurable deviations should exist between both the considered theories. We performed corresponding measurements, the results of which were published in [20,21].

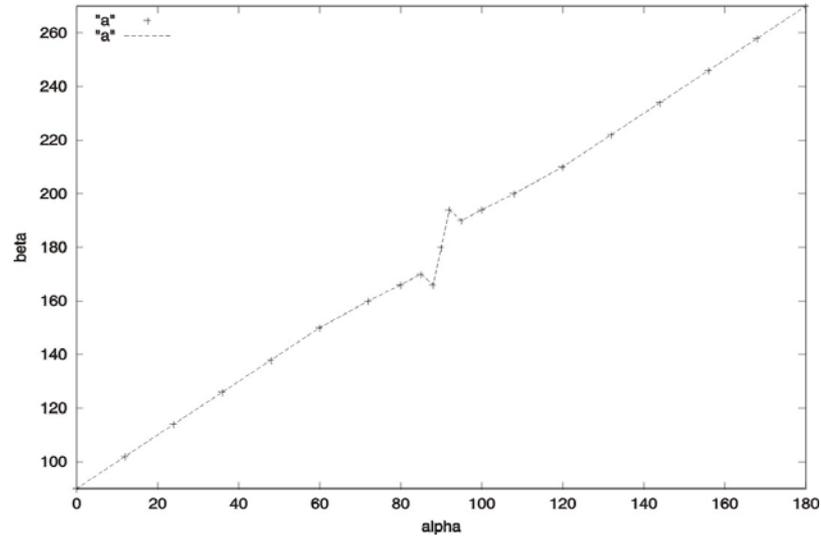

Fig. 1:  The pairs of angles α  and  β (used for the measurement shown in Fig. 2)

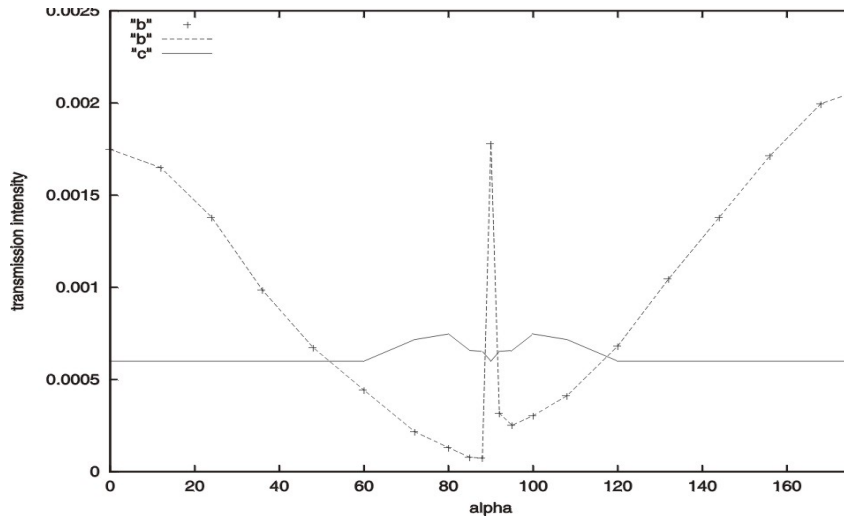

Fig. 2:  Light transmission through three polarizers (for  α - β pairs shown in Fig.1);
experimental data - dashed line; quantum-mechanical (orthodox) prediction – full line.

The experimental results (see [21]) obtained for angle combinations, shown in Fig. 1 in the present paper, are represented by dashed line in Fig. 2 (see also [22]). For a given angle  α  the angle  *β*  was always established  so as the total light transmission was minimum. The quantum-mechanical prediction for these angle pairs is represented by full line; the position of this line being shifted in vertical direction somewhat arbitrarily as the values of parameter  ε  for given "imperfect" polarizers were not available. In any case the quantum-mechanical prediction gives maximums at the positions where the experiment exhibits deep minimums. The given discrepancy was not accented in [21] as we were afraid reasonably that the paper would not have been accepted for publication in such a case.

### 9. Extended quantum-theoretical mathematical model and metaphysics

We have already mentioned that the modern physics was significantly influenced by philosophical opinions that ruled in Europe in the end of the 19$^{th}$ and in the beginning of the 20$^{th}$ centuries. However, at the present the situation has changed and the postmodern philosophy has referred often to the modern physics (and mainly to Copenhagen quantum mechanics) as to the most important support for its statements. The physicists take then the basic assumptions of physical theories as fixed laws and do not take any care at all that the knowledge develops and that it is necessary to continue also in falsification efforts. Instead of it they discard rather any parallel hypotheses and theories from further considerations on the basis of falsifiability ideas and sometimes also experimental data that contradict standard theories. An example may be represented by the mentioned data obtained with three polarizers [20,21] that were published more then 10 years ago and have never been doubted; only their existence has not been taken into account.

In the preceding sections sufficient amount of arguments has been provided that some statements used permanently as the support of earlier assumptions and theoretical approaches are evidently wrong. It has been also shown that the extended (generalized) mathematical model (described in papers [5,6]) has removed all known critical standpoints concerning the quantum mechanics and may be denoted in the sense of falsification approaches as plausible for the contemporary period of our knowledge; statistical (ensemble) interpretation being strongly preferred. A decisive change against the Copenhagen alternative consists in that the evolution of physical systems may be again described as causal and practically deterministic. The mathematical model as the whole remains, however, without any greater change. It is only necessary to modify the second and to abandon the third assumption (see Sec. 4) that have changed the content of Schroedinger equation, which has led to known paradoxical properties.

There are, of course, random phenomena in the nature. However, it is the chance as it was defined already by Aristotle, i.e., as the meeting (interaction) of two causally developing chains that cannot be correlated at all before. It may be done only according to consequences that followed after their interaction. The chance cannot be in any case a substantial part of each step in the time evolution. It is necessary to go back to causality, which means that no support for postmodern thinking may follow from physical knowledge. Our physical results contradict it rather.

It follows also from the preceding that there is not any reason for abandoning the two-value logic proposed by Aristotle. Any falsification does not take place in this direction. And there is not any reasonable sign, either, that would tend to such possibility.

As the more general quantum-mechanical mathematical model based on the extended Hilbert space has opened again the way to realistic description of microworld as well as of macroworld, there is not quite any reason for limiting to phenomenological models, even if they may be useful in some cases. The discussed more general model is based fully on ontological approach and requires to take into account the whole reality of physical being and acting, and not only a chosen set of measured values. It is also necessary to finish with deforming trends that have followed from quite unsubstantial and mistaken interpretations having been derived oft from phenomenological models.

### 10. Conclusion

The preceding text has summarized the recent results concerning some important aspects of quantum mechanics. It has been shown that the theoretical as well as experimental results have led to the conclusion that the statistical interpretation (being practically equivalent to hidden-variable theory) should be strongly preferred. The critical comments of mathematical as well as physical kinds (discussed permanently in literature) have been summarized and discussed. And it has been shown that practically all may be removed if the basis of assumptions the Copenhagen quantum mechanics has started from has been modified. The results of experiments with three polarizers have been also introduced that are according to our opinion in decisive conflict with standardly applied quantum mechanics.

The role has been also critically examined, which has been played by different kinds of physical models representing an indivisible part of natural science at the present. The paper contains also the attempt of a natural scientist who was interested during his whole life in corresponding interdisciplinary questions to express his opinion to relation between the physical science and philosophical (metaphysical) tendencies and how they influenced mutually in the course of the last centuries. Especially, it has been examined critically what role was played in this process by different kinds of physical models that have represented an indivisible part of natural science.

All approaches and physical results are fully documented in quoted papers, the part of which has been published in standard journals and the other part being available on Internet. In the presented paper mainly the results and conclusions have been summarized that are important for the discussion of the relation between physical concepts and philosophical (metaphysical) point of view.

It follows also from the preceding that it will be necessary to take a different standpoint to scientific knowledge. It is not possible to separate fully the science and metaphysics, as the system of human knowledge cannot be formulated if the logical rules do not belong to basic hypotheses. Similarly, our knowledge expressed by positive statements cannot be denoted as verified, but always only as plausible if one did not succeed in falsifying them. Only the set of all negative statements that have been derived on the basis of falsification approaches may be denoted as the certain truth.

**Appendix  -  Several words about author life lot**

This text would remain incomplete and to some extent incomprehensible if at least several additional words were not said about the life lot of its author; the paper could be hardly written without this external intervention. He started his carrier in physical research in the end of the fortieth years of the past century. In the beginning of the fiftieth years he was co-author of paper [23], in which it was shown for the first time that the idea of isotopic spin might be useful in classifying at least some elementary particles having been newly discovered at that time. At the present this phenomenological quantity represents a part of all corresponding physical tables.

The then physical approaches were based on phenomenological attitude. And the author would have continued also in further research activities undoubtedly in this direction if his life had not been influenced by decisions of communist governors in his country. He was arrested and put in prison; after getting again free he was not allowed to go back to previous work. However, the fate chose for him a new path. Quite unexpectedly, he succeeded in 1959 in getting the job in the Research Enterprise Tesla-Přemyšlení where his duty was to take part in preparation of newly developed detectors and measuring devices of ionizing radiation of very low energies and very low activities for medical purposes. He had to work fully in experimental and technical region, which did not happen probably to any theoretical physicist with similar qualification. He was forced to deal with actual behavior of individual elementary particles (including photons) which were earlier main subjects of his theoretical interest.

It was evident soon that his qualified knowledge of the then theoretical ideas was not at all useful. The nature showed to consist of concrete matter objects with given properties, for which the propagated phenomenological theoretical descriptions (based on field theories) were quite inapplicable. That represented for him a decisive impulse to new considerations that would be based on ontological grounds. He started to work out new ideas since 1968 when under somewhat changed political conditions he succeeded in returning back to original scientific activities.

He began to work in new direction parallel with other tasks and duties. He was meeting permanently with opposition of colleagues following from the ideology of falsifiability. It was possible to publish only a part of finished (more mathematically oriented) papers in corresponding journals. All other necessary papers are available on Internet (see attached quotations). The problems consisted partially also in that the newly applied ontological access involved the whole physical basis. And it was very difficult to explain sufficiently all concerned aspects in one shorter paper, when each reader started from the knowledge and conviction gained in the past. In addition to, a great amount of critical comments was directed to the problems lying aside and not representing proper subjects of the presented results. Thus, for the author also new questions permanently emerged that should have been answered in the framework of the ontological concept.

It seems, however, that at the present the solution of corresponding set of related problems may be denoted as closed. It is possible to say that a sufficient amount of new results has been gathered and the time has come when it is necessary (and also fully possible) to push the whole problem of ontological models to a new stage.

e-mail: lokaj @ fzu.cz   (August  2006)